\documentclass{article}
\usepackage{spconf,amsmath,graphicx}
\usepackage{algorithm2e}
\usepackage{amsfonts}
\usepackage{url}
\usepackage{booktabs}  
\usepackage{multirow}  
\usepackage{array}     
\usepackage{rotating}  

\title{Image Motion Blur Removal in the Temporal Dimension with Video Diffusion Models}
\name{Wang Pang$^{\star}$$^{\dagger}$ \thanks{$^{\star}$: These authors contributed equally.} \qquad Zhihao Zhan$^{\star}$$^{\ddagger}$ \qquad Xiang Zhu$^{\star}$$^{\ddagger}$ \qquad Yechao Bai$^{\dagger}$}
  
\address{$^{\dagger}$ Nanjing University \qquad
      $^{\ddagger}$ TopXGun Robotics}
\begin{document}
%
\maketitle
\begin{abstract}
Most motion deblurring algorithms rely on spatial-domain convolution models, which struggle with the complex, non-linear blur arising from camera shake and object motion. In contrast, we propose a novel single-image deblurring approach that treats motion blur as a temporal averaging phenomenon. Our core innovation lies in leveraging a pre-trained video diffusion transformer model to capture diverse motion dynamics within a latent space. It sidesteps explicit kernel estimation and effectively accommodates diverse motion patterns. We implement the algorithm within a diffusion-based inverse problem framework. Empirical results on synthetic and real-world datasets demonstrate that our method outperforms existing techniques in deblurring complex motion blur scenarios. This work paves the way for utilizing powerful video diffusion models to address single-image deblurring challenges.
\end{abstract}
\section{Introduction}
\label{sec:intro}

Motion of the camera or objects during the exposure time leads to motion blur, which is very common in imaging processes~\cite{dai2008motion}. Removing such blur is never trivial. In the past two decades, numerous algorithms have been proposed for motion deblurring (MD), and they are generally categorized into two types: those with explicit kernel estimation and those without.

Kernel-based methods assume that motion blur can be approximated by a convolution model~\cite{dai2008motion, shan2008high, anger2019efficient, kaufman2020deblurring}:
\begin{equation}
  \mathbf{y} = \mathbf{x} \ast \mathbf{h} + \mathbf{e},
  \label{eq:conv_model}
\end{equation}
where $\mathbf{x}$ is the underlying sharp image, $\mathbf{h}$ is an unknown blur kernel representing the motion trajectory of either the camera or objects, $\ast$ denotes the convolution operator, and $\mathbf{e}$ is the additive Gaussian sensing noise. Such methods typically estimate both $\mathbf{x}$ and $\mathbf{h}$ using maximum a posteriori (MAP) estimations~\cite{shan2008high, anger2019efficient} or convolutional neural networks (CNNs)~\cite{kaufman2020deblurring}. However, spatially varying blur due to camera or object motion makes precise kernel estimation nearly impossible. Furthermore, real-world object movement can involve complex, non-linear trajectories that cannot be captured by simple convolution, even when layer decomposition strategies are applied~\cite{dai2008motion}. Consequently, the convolution assumption rarely holds in practice, limiting the effectiveness of kernel-based approaches.

Kernel-free methods are mainly based on deep neural networks. They could be CNNs~\cite{zhang2023event}, RNNs~\cite{park2020multi}, or Transformers~\cite{liang2024image}. Most of them are trained via supervised learning, and some use GANs~\cite{kupyn2018deblurgan}. In general, they assume there is a one-to-one mapping between the underlying sharp image $\mathbf{x}$ and the observed blurry image $\mathbf{y}$, and as long as there is sufficient such image data (paired or unpaired), a neural network $F_{\mathbf{\theta}}(\cdot)$ can be employed, with its weights $\theta$ tuned to learn the mapping from a blurred image to its sharp version:
\begin{equation}
  \mathbf{x} = F_{\mathbf{\theta}}(\mathbf{y}).
  \label{eq:nn_solution}
\end{equation}

However, the mathematical relationship between the blurred-sharp image pair remains ambiguous. In fact, some research on synthesizing image pairs for motion deblurring model training shows that the synthesizing process is not a one-to-one mapping but rather an \(N\)-to-one mapping~\cite{nah2017deep, wu2024id}:
\begin{equation}
    \mathbf{y} = \frac{1}{P}\int_{0}^{P}\mathbf{x}(\tau)\mathrm{d}\tau + \mathbf{e} .
    \label{eq:integral_model}
\end{equation}
Here, images are assumed to be in linear color space. $P$ represents exposure time, and $\mathbf{x}(\tau)$ denotes the ideal sharp image taken at time $\tau$ with an infinitely short exposure time. In practice, this integral process can be approximated by averaging over \(N\) sharp frames $\{\mathbf{x}_n\}$ taken by a high-speed camera~\cite{wu2024id}:

\begin{equation}
    \mathbf{y} \approx
    \frac{1}{N}\sum_{n=0}^{N-1}\mathbf{x}_n + \mathbf{e}.
    \label{eq:average_model}
\end{equation}

Compared with convolution in the spatial domain, this temporal averaging model is more natural and much simpler. It does not require any kernel estimation or foreground-background segmentation, and \(N\) can be easily estimated via the actual exposure time. So, why have we not yet seen deblurring algorithms based on this model? The answer lies in its highly ill-posed nature. Strong prior knowledge about the video \(\{\mathbf{x}_n\}\) is required to estimate it from a single-frame observation \(\mathbf{y}\).

In this paper, we argue that an unconditional video diffusion model (VDM), which learns the prior distribution of not only the image contents but also the object movements, can be used to estimate \(\{\mathbf{x}_n\}\) in a latent space from a given blurry \(\mathbf{y}\). The estimation is performed by solving an inverse problem under the Diffusion Posterior Sampling (DPS) framework~\cite{chung2022diffusion}.

The original DPS has already shown its robust reconstruction capabilities in blind global motion deblurring~\cite{chung2022diffusion, chung2023parallel}, but that is still based on the convolution model Eq.~\eqref{eq:conv_model}. Our algorithm introduces several novelties:

\begin{itemize}
\item It processes 3D videos in their entirety, learning not only the distribution of their visual contents but also the dynamics governed by real-world physics.
\item It employs a transformer network, rather than a UNet, as the denoiser in the reverse diffusion process, enhancing the model’s ability to scale and handle complex, dynamic scenarios more effectively.
\item It manages visual statistics within a latent space to reduce dimensionality.
\item It uses Eq.~\eqref{eq:average_model} as the degradation model without any kernel estimation and can handle various motions in principle. Its output is deblurred video frames instead of a single sharp image.
\end{itemize}

To validate our approach, we conduct experiments on synthetic video datasets to analyze its behavior and evaluate its performance on real-world videos.

\section{Related work}
\label{sec:related}

\subsection{Diffusion Models}
Diffusion models have recently shown remarkable success in generating multi-dimensional signals such as images, videos and audios. The core idea is to learn the prior distribution of data $\mathbf{x}$ by gradually adding Gaussian noise to a clean sample until it becomes pure noise, then training a network to reverse this noising process step by step. Formally, the forward noising can be represented by a stochastic differential equation (SDE)~\cite{song2020score}:
\begin{equation}
  d\mathbf{x} = -\frac{\beta(t)}{2}\mathbf{x}_t dt + \sqrt{\beta(t)}d\mathbf{w},
  \label{eq:sde_forward2}
\end{equation}
where $\beta(t)$ is the noise schedule, $\mathbf{w}$ denotes a standard Brownian motion, and $d\mathbf{w}$ represents white Gaussian noise. Reversing this process involves:
\begin{equation}
  d\mathbf{x} = \left(-\frac{\beta(t)}{2}\mathbf{x}_t - \beta(t)\nabla_{\mathbf{x}_t} \log p(\mathbf{x}_t) \right)dt + \sqrt{\beta(t)}d\mathbf{w},
  \label{eq:sde_reverse1}
\end{equation}
where $\nabla_{\mathbf{x}_t} \log p(\mathbf{x}_t)$ is the score function of the unknown distribution $p(\mathbf{x}_t)$. This score function can be approximated by a neural network $\mathbf{s}_{\theta}(\mathbf{x}_{t}, t)$ via score matching:
\begin{equation}
  \theta^{*} = \arg\min_{\theta}\mathbb{E}_{t, \mathbf{x}_t, \mathbf{x}_0}\left( \| \nabla_{\mathbf{x}_t} \log p(\mathbf{x}_t|\mathbf{x}_0) - \mathbf{s}_{\theta}(\mathbf{x}_{t}, t) \|_{2}^{2} \right),
  \label{eq:score_matching}
\end{equation}
This learned network replaces the score function in Eq.~\eqref{eq:sde_reverse1}, enabling incremental denoising from pure noise back to a sample drawn from the underlying data distribution.

\subsection{Diffusion Posterior Sampling (DPS)}
Chung et al.~\cite{chung2022diffusion} extended diffusion models to solve inverse problems, such as deconvolution and super-resolution, by introducing the DPS framework. 

Again, let $\mathbf{x}$ be the ideal data vector and $\mathbf{y}$ the lower-dimensional or degraded observation. Assuming a known degradation operator $H(\cdot)$ with additive noise $\mathbf{e} \sim \mathcal{N}(0, \sigma^{2})$, we have
\begin{equation}
  \mathbf{y} = H(\mathbf{x}) + \mathbf{e}, 
  \quad
  p(\mathbf{y} | \mathbf{x}) = \mathcal{N}(\mathbf{y} | H(\mathbf{x}), \sigma^{2}\mathbf{I} ).
  \label{eq:observation}
\end{equation}
Combining the prior $p(\mathbf{x})$ and likelihood $p(\mathbf{y}|\mathbf{x})$ via Bayes' rule yields the conditional score:
\begin{equation}
  \nabla_{\mathbf{x}} \log p(\mathbf{x} | \mathbf{y}) = 
  \nabla_{\mathbf{x}} \log p(\mathbf{y} | \mathbf{x}) + \nabla_{\mathbf{x}} \log p(\mathbf{x}).
  \label{eq:posterior_score}
\end{equation}
DPS incorporates this conditional score into the reverse diffusion process by approximating the likelihood term at each denoising step. Specifically,
\begin{equation}
  \nabla_{\mathbf{x}_t} \log p(\mathbf{y} | \mathbf{x}_t) \approx 
  \nabla_{\mathbf{x}_t} \log p(\mathbf{y} | \hat{\mathbf{x}}_0(\mathbf{x}_t)),
  \label{eq:dps_approx1}
\end{equation}
where
\begin{equation}
  \hat{\mathbf{x}}_0(\mathbf{x}_t) = \frac{1}{\sqrt{\bar{\alpha}(t)}}\left(\mathbf{x}_t + (1 - \bar{\alpha}(t))\mathbf{s}_{\theta^*}(\mathbf{x}_t, t)\right).
  \label{eq:dps_approx2}
\end{equation}
Hence, the reverse SDE from Eq.~\eqref{eq:sde_reverse1} is modified to include the observation model:
\begin{equation}
\begin{split}
  d\mathbf{x} = & \biggl[-\frac{\beta(t)}{2}\mathbf{x}_t - \beta(t)\Bigl( \mathbf{s}_{\theta^*}(\mathbf{x}_t, t) \\
  & - \frac{1}{\sigma^2} \nabla_{\mathbf{x}_t} \|\mathbf{y} - H(\hat{\mathbf{x}}_0(\mathbf{x}_t)) \| \Bigr) \biggr] dt + \sqrt{\beta(t)}d\mathbf{w}.
  \label{eq:dps_reverse}
\end{split}
\end{equation}
where the learned prior and observation likelihood jointly refine $\mathbf{x}_t$ at each iteration.

\label{sec:approach}
\begin{figure*}
    \centering
    \includegraphics[width=1.0\linewidth]{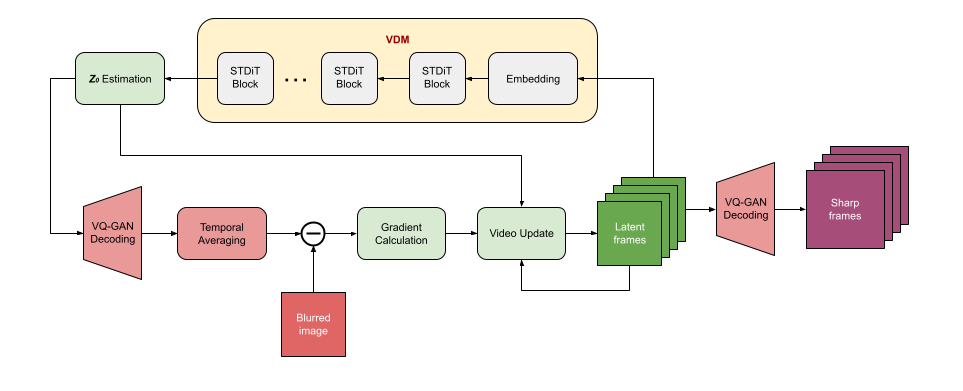}
    \caption{Overview of the VDM-MD method: In the core iteration, the estimated 3D sharp video resides in the latent space, represented by green boxes. It is generated and refined by the pre-trained VDM, which includes several STDiT blocks. The latent video is then decoded and compared with the blurry image through the degradation model, indicated by red boxes. Their discrepancies are used to correct and enhance the video. Upon completion the latent video is decoded back to the visual space.}
    \label{fig:overview}
\end{figure*}

\subsection{Video Diffusion Models (VDMs)}
Recent VDMs have yielded strikingly realistic results in video generation~\cite{gupta2023photorealistic, openai2024sora, zheng2024open}, often by employing transformer-based architectures with strong scalability and parallelization. Many of these methods use diffusion transformers. A notable example is Sora~\cite{openai2024sora}, which demonstrates two key strengths: temporal coherence across frames and realistic object movements that closely mimic real-world physics. Such capabilities suggest an intriguing possibility: if VDMs can accurately track complex motions, might they also serve as effective “world models” for single-image motion deblurring when cast as an inverse problem?

\section{Proposed Approach}
\label{sec:proposed}

We present \textbf{VDM-MD}, a VDM based method that formulates motion deblurring as an inverse problem within the DPS framework. Our key premise is that once the VDM has learned the underlying dynamics of a world represented by a training video dataset, it can naturally resolve single image motion blur as long as the image is about the given world. An overview of the architecture is shown in Figure~\ref{fig:overview}.

We adopt the temporal averaging model from Eq.~\eqref{eq:average_model}, expressed as:
\begin{equation}
  \mathbf{y} = H(\mathbf{X}) + \mathbf{e},
  \label{eq:frame_degrade_2}
\end{equation}
where $\mathbf{X} \in \mathbb{R}^{N \times H \times W \times 3}$ represents the ideal sharp video frames, $\mathbf{y} \in \mathbb{R}^{H \times W \times 3}$ denotes the observed motion blurred image, and $\mathbf{e}$ is white Gaussian noise with covariance $\sigma^2 \mathbf{I}$.

Similar to DPS, the prior distribution of $\mathbf{X}$ is defined by a pre-trained diffusion model. However, unlike the original DPS, we perform diffusion sampling in a latent space learned by a VQ-GAN~\cite{esser2021taming} to handle high-dimensional video data efficiently. We remove the quantization step and apply VQ-GAN only spatially with a compression factor of $p=8$. This transforms $\mathbf{X}$ into a latent tensor $\mathbf{Z} \in \mathbb{R}^{N \times (H/p) \times (W/p) \times c}$, where $c$ is the number of latent channels. The decoder $D(\cdot)$ then reconstructs $\mathbf{X}$ from $\mathbf{Z}$.

Given a latent video $\mathbf{Z}$, the conditional likelihood of the observed blurry image $\mathbf{y}$ is:
\begin{equation}
  p(\mathbf{y} | \mathbf{Z}) = \mathcal{N}(\mathbf{y} | H(D(\mathbf{Z})), \sigma^2 \mathbf{I}).
  \label{eq:new_conditional_prob}
\end{equation}
where $H(D(\cdot))$ remains differentiable, thus allowing integration into the DPS framework. The corresponding reverse diffusion equation becomes:
\begin{equation}
\begin{split}
  d\mathbf{Z} = & \biggl[-\frac{\beta(t)}{2}\mathbf{Z}_t - \beta\Bigl( \mathbf{s}_{\theta^*}(\mathbf{Z}_t, t) \\
  & - \frac{1}{\sigma^2} \nabla_{\mathbf{Z}_t} \|\mathbf{y} - H(D(\hat{\mathbf{Z}}_0(\mathbf{Z}_t))) \| \Bigr) \biggr] dt \\
  & + \sqrt{\beta(t)}d\mathbf{W}.
  \label{eq:our_dps_reverse}
\end{split}
\end{equation}
where $\mathbf{s}_{\theta^*}(\mathbf{Z}, t)$ is an unconditional diffusion network trained in the latent video space. In our case, we utilize a DiT-based architecture similar to the STDiT model from Open-Sora~\cite{zheng2024open}, but without conditional embeddings. The complete reverse process is summarized in Algorithm~\ref{alg:main}.

\begin{figure*}[ht]
    \centering
    \begin{minipage}[c]{0.20\linewidth}
        \centering
        \begin{tabular}{@{}m{0.03\linewidth}<{\centering} m{0.90\linewidth}@{}}
            \rotatebox{90}{\textbf{Blurry input 17326}} &
            \includegraphics[width=\linewidth]{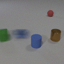}
        \end{tabular}
    \end{minipage}
    \hspace{0.02\linewidth}
    \begin{minipage}[c]{0.50\linewidth}
        \centering
        
        \begin{tabular}{@{}m{0.01\linewidth}<{\centering} m{0.95\linewidth}@{}}
            \rotatebox{90}{\textbf{GT}} &
            \begin{minipage}[c]{\linewidth}
                \centering
                \includegraphics[width=0.24\linewidth]{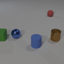}%
                \hfill
                \includegraphics[width=0.24\linewidth]{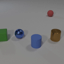}%
                \hfill
                \includegraphics[width=0.24\linewidth]{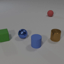}%
                \hfill
                \includegraphics[width=0.24\linewidth]{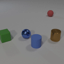}%
            \end{minipage}
        \end{tabular}

        \vspace{0.5em} 

        \begin{tabular}{@{}m{0.01\linewidth}<{\centering} m{0.95\linewidth}@{}}
            \rotatebox{90}{\textbf{Output}} &
            \begin{minipage}[c]{\linewidth}
                \centering
                \includegraphics[width=0.24\linewidth]{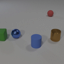}%
                \hfill
                \includegraphics[width=0.24\linewidth]{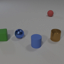}%
                \hfill
                \includegraphics[width=0.24\linewidth]{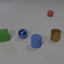}%
                \hfill
                \includegraphics[width=0.24\linewidth]{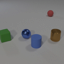}%
            \end{minipage}
        \end{tabular}
    \end{minipage}

    \vspace{1em} 

    \begin{minipage}[c]{0.20\linewidth}
        \centering
        \begin{tabular}{@{}m{0.03\linewidth}<{\centering} m{0.90\linewidth}@{}}
            \rotatebox{90}{\textbf{Blurry input 17032}} &
            \includegraphics[width=\linewidth]{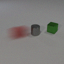}
        \end{tabular}
    \end{minipage}
    \hspace{0.02\linewidth}
    \begin{minipage}[c]{0.50\linewidth}
        \centering
        
        \begin{tabular}{@{}m{0.01\linewidth}<{\centering} m{0.95\linewidth}@{}}
            \rotatebox{90}{\textbf{GT}} &
            \begin{minipage}[c]{\linewidth}
                \centering
                \includegraphics[width=0.24\linewidth]{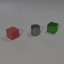}%
                \hfill
                \includegraphics[width=0.24\linewidth]{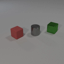}%
                \hfill
                \includegraphics[width=0.24\linewidth]{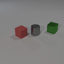}%
                \hfill
                \includegraphics[width=0.24\linewidth]{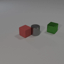}%
            \end{minipage}
        \end{tabular}

        \vspace{0.5em} 

        \begin{tabular}{@{}m{0.01\linewidth}<{\centering} m{0.95\linewidth}@{}}
            \rotatebox{90}{\textbf{Output}} &
            \begin{minipage}[c]{\linewidth}
                \centering
                \includegraphics[width=0.24\linewidth]{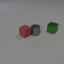}%
                \hfill
                \includegraphics[width=0.24\linewidth]{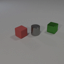}%
                \hfill
                \includegraphics[width=0.24\linewidth]{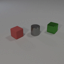}%
                \hfill
                \includegraphics[width=0.24\linewidth]{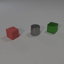}%
            \end{minipage}
        \end{tabular}
    \end{minipage}

    \caption{Motion deblurring examples with CLEVRER dataset. Each blurry inputs are generated by averaging 10 frames. Only the 0th, 3rd, 6th, and 9th frame of the GT and output videos are illustrated.}
    \label{fig:clevrer_visual_1}
\end{figure*}

\section{Experiments}
\label{sec:experiment}

\subsection{Synthetic Dataset}
To analyze our algorithm’s performance without requiring an extensive, large-scale transformer, we used the CLEVRER dataset~\cite{srivastava2015unsupervised} as a “toy world.” CLEVRER features relatively simple objects obeying basic physics, with minimal motion between consecutive frames. Each video clip thus approximates a high-frame-rate recording.

We extracted 50k clips at a resolution of $10 \times 64 \times 64 \times 3$ for training and synthesized blurry images by averaging the 10 frames of each test clip. After VQ-GAN compression the dimension of each video clip is reduced to $10 \times 8 \times 8 \times 12$. Our VDM contains 28 STDiT layers and 726M parameters, and it was trained on 4 4090 GPUs.

Figure~\ref{fig:clevrer_visual_1} shows representative results alongside ground truth scenes. For example, \emph{sample 17326}, featuring translation and self-rotation, is nearly perfectly reconstructed; \emph{sample 17032} appears reversed in time, reflecting Newtonian time-reversible dynamics in CLEVRER (where a single blurred image lacks directional cues).

To assess robustness against mismatches between our assumed frame-averaging model $H(\cdot)$ and real-world blur formation, we introduced a temporal down-sampling experiment. Starting with 40 frames indexed $\{0, 1, \dots, 39\}$, we retained only every 4th frame $\{0, 4, 8, \dots, 36\}$ for training. During testing, we produced two types of blurry images:
\begin{itemize}
  \item \emph{Smoothly Blurred Images}, averaging all 37 original frames $\{0, 1, \dots, 36\}$;
  \item \emph{Less Smoothly Blurred Images}, averaging only the 10 down-sampled frames $\{0, 4, 8, \dots, 36\}$.
\end{itemize}

This deliberate mismatch simulates the gap between our temporal averaging model~\eqref{eq:average_model} and the real integral process~\eqref{eq:integral_model}. Despite the difference in frame rates, the final deblurring performance remained nearly unchanged: our method consistently recovered high-fidelity sharp frames with minimal visual artifacts, and PSNR/SSIM metrics (over 500 test videos) varied only slightly between smoothly and less smoothly blurred inputs (see Table~\ref{tab:psnr_ssim_two_methods}). These findings indicate that while $H(\cdot)$ may not perfectly match real-motion conditions, the learned video diffusion model is robust to such deviations. It also suggests there is no strong need for training with high-speed camera data in practice.

\begin{figure*}[ht]
    \centering
    \begin{minipage}[c]{0.155\textwidth} 
        \centering
        \includegraphics[width=0.9\linewidth]{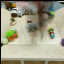}\\
    \end{minipage}
    \vspace{0.5em}
    \begin{minipage}[c]{0.155\textwidth}
        \centering
        \includegraphics[width=0.9\linewidth]{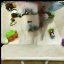}\\
    \end{minipage}
    \vspace{0.5em}
    \begin{minipage}[c]{0.155\textwidth}
        \centering
        \includegraphics[width=0.9\linewidth]{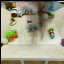}\\
    \end{minipage}
    \vspace{0.5em}
    \begin{minipage}[c]{0.155\textwidth}
        \centering
        \includegraphics[width=0.9\linewidth]{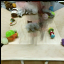}\\
    \end{minipage}
    \vspace{0.5em}
    \begin{minipage}[c]{0.155\textwidth}
        \centering
        \includegraphics[width=0.9\linewidth]{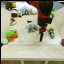}\\
    \end{minipage}
    \vspace{0.5em}
    \begin{minipage}[c]{0.155\textwidth}
        \centering
        \includegraphics[width=0.9\linewidth]{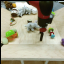}\\
    \end{minipage}

    \vspace{-2em} 

    \begin{minipage}[c]{0.155\textwidth}
        \centering
        \includegraphics[width=0.9\linewidth]{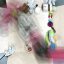}\\
    \end{minipage}
    \vspace{0.5em}
    \begin{minipage}[c]{0.155\textwidth}
        \centering
        \includegraphics[width=0.9\linewidth]{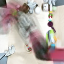}\\
    \end{minipage}
    \vspace{0.5em}
    \begin{minipage}[c]{0.155\textwidth}
        \centering
        \includegraphics[width=0.9\linewidth]{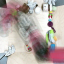}\\
    \end{minipage}
    \vspace{0.5em}
    \begin{minipage}[c]{0.155\textwidth}
        \centering
        \includegraphics[width=0.9\linewidth]{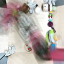}\\
    \end{minipage}
    \vspace{0.5em}
    \begin{minipage}[c]{0.155\textwidth}
        \centering
        \includegraphics[width=0.9\linewidth]{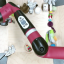}\\
    \end{minipage}
    \vspace{0.5em}
    \begin{minipage}[c]{0.155\textwidth}
        \centering
        \includegraphics[width=0.9\linewidth]{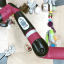}\\
    \end{minipage}

    \vspace{-2em} 

    \parbox{0.155\textwidth}{\centering \small Blurry}%
    \vspace{0.5em}
    \parbox{0.155\textwidth}{\centering \small MPRNet~\cite{zamir2021multi}}%
    \vspace{0.5em}
    \parbox{0.155\textwidth}{\centering \small MTRNN~\cite{park2020multi}}%
    \vspace{0.5em}
    \parbox{0.155\textwidth}{\centering \small Restormer~\cite{zamir2022restormer}}%
    \vspace{0.5em}
    \parbox{0.155\textwidth}{\centering \small VDM-MD (5th frame)}%
    \vspace{0.5em}
    \parbox{0.155\textwidth}{\centering \small GT (5th frame)}%
    
    \vspace{-3em} 
    
    \caption{Comparison on BAIR dataset. For the GT and reconstructed videos only the 5th (middle) frame is shown. }
    \label{fig:bair_comparison}
\end{figure*}

\begin{table}[ht]
    \centering
    \caption{Comparison between Two Type of Inputs.}
    \label{tab:psnr_ssim_two_methods}
    \begin{tabular}{lcc}
        \toprule
        \textbf{Input Type} & \textbf{PSNR} & \textbf{SSIM} \\
        \midrule
        \textbf{Smoothly Blurred} & 30.26 & 0.914 \\
        \textbf{Less Smoothly Blurred} & 29.93 & 0.911 \\
        \bottomrule
    \end{tabular}
\end{table}

\subsection{BAIR Dataset}
To evaluate our method on real-world data, we used the BAIR robot pushing dataset~\cite{ebert2017self}, which consists of 90K short video clips recorded by a real camera. Although this setting remains somewhat of a “toy world” (featuring robotic arms in a controlled environment), it introduces more natural lighting, scene textures, and frequent occlusions than CLEVRER.

Because the dataset does not include truly motion-blurred images, we synthesized blurred inputs by averaging consecutive frames, similar to our CLEVRER setup. We trained our model using 260K video clips. The other settings are the same as the CLEVRER tests. Our approach effectively reconstructed sharp videos under these conditions, often achieving near-perfect restoration of the robot arm’s position and scene details (see Figure~\ref{fig:bair_comparison}). In some cases, motion is reversed in time due to the inherent ambiguity of single-image blur.

We compared our algorithm to three state-of-the-art single-image deblurring methods: MPRNet~\cite{zamir2021multi}, MTRNN~\cite{park2020multi}, and Restormer~\cite{zamir2022restormer}, each producing only a single deblurred image. Because there is no exact single-frame ground truth for each blurred observation, we took the 5th (middle) frame of our recovered sequence for quantitative evaluation, then measured PSNR and SSIM against the corresponding middle ground-truth frame (see Table~\ref{tab:psnr_ssim}). All three baselines struggled to remove the local motion blur caused by complex robot movements, which is unsurprising given they were never designed to handle this kind of motion-blur scenario. This result highlights the advantage of treating blurred content as a short video rather than relying on a single-sharp-image assumption.

\begin{table}[ht]
    \centering
    \caption{Quantitative Comparison on BAIR dataset.}
    \label{tab:psnr_ssim}
    \begin{tabular}{l cc cc}
        \toprule
        & \multicolumn{2}{c}{\textbf{BAIR Main}} 
        & \multicolumn{2}{c}{\textbf{BAIR Aux1}} \\
        \cmidrule(r){2-3} 
        \cmidrule(l){4-5}
        \textbf{Method} & \textbf{PSNR} & \textbf{SSIM} & \textbf{PSNR} & \textbf{SSIM} \\
        \midrule
        MPRNet~\cite{zamir2021multi} & 16.45 & 0.718 & 19.30 & 0.789 \\
        MTRNN~\cite{park2020multi} & 16.57 & 0.723 & 21.29 & 0.867 \\
        Restormer~\cite{zamir2022restormer} & 16.58 & 0.728 & 19.95 & 0.818 \\
        \textbf{VDM-MD} & \textbf{24.24} & \textbf{0.896} & \textbf{26.78} & \textbf{0.948} \\
        \bottomrule
    \end{tabular}
\end{table}

\section{Conclusions}
\label{sec:conclude}
\RestyleAlgo{ruled}
\begin{algorithm}
\caption{VDM-MD}\label{alg:main}
\KwIn{$\mathbf{y}$, $T$}
Initialize $\mathbf{Z}_T \sim \mathcal{N}(\mathbf{0}, \mathbf{I})$\;
\For{$t \leftarrow T - 1$ \KwTo $0$}{
    $\hat{\mathbf{s}} = \mathbf{s}_{\theta^*}(\mathbf{Z}_t, t)$\;
    $\hat{\mathbf{Z}}_0 = \dfrac{1}{\sqrt{\bar{\alpha}_t}} \left(\mathbf{Z}_t + (1 - \bar{\alpha}_t)\hat{\mathbf{s}}\right)$\;
    Sample $\boldsymbol{\epsilon} \sim \mathcal{N}(\mathbf{0}, \mathbf{I})$\;
    $\mathbf{Z}'_{t-1} = \dfrac{\sqrt{\alpha_t} (1 - \bar{\alpha}_{t-1})}{1 - \bar{\alpha}_t}\mathbf{Z}_t + \dfrac{\sqrt{\bar{\alpha}_{t-1}}\beta_t}{1 - \bar{\alpha}_t}\hat{\mathbf{Z}}_0 + \sigma_t\boldsymbol{\epsilon}$\;
    $\hat{\mathbf{y}}_{t-1} = H\left(D(\hat{\mathbf{Z}}_0)\right)$\;
    $\mathbf{Z}_{t-1} = \mathbf{Z}'_{t-1} - \eta_t \nabla_{\mathbf{Z}_t}\left\|\mathbf{y} - \hat{\mathbf{y}}_{t-1} \right\|_2^2$\;
}
$\hat{\mathbf{X}} = D(\hat{\mathbf{Z}}_0)$\;
\KwOut{$\hat{\mathbf{X}}$}
\end{algorithm}
We introduced a single image motion deblurring approach that reinterprets the task as a video diffusion problem, recovering multiple sharp frames instead of a single deblurred image. Central to our method is the ability to learn not only the distribution of visual content, but also the underlying physics that govern motion in real-world scenes. By employing a transformer network in the diffusion process, our system scales effectively to complex, dynamic scenarios, while managing high-dimensional video data in a latent space to reduce computational overhead. It forgoes explicit kernel estimation by adopting a temporal averaging model, thus accommodating a wide range of motion patterns.

Despite these advantages, our current setup cannot yet serve as a fully general-purpose solution, primarily due to limited computational resources and training data. Real-world deployment would require a large-scale diffusion model, such as commercial platforms like OpenAI’s Sora or Google’s Veo. Nonetheless, our findings demonstrate the potential of leveraging powerful video diffusion models for single-image deblurring and highlight a promising direction for both academic research and industrial applications. 

\vfill\pagebreak

\bibliographystyle{IEEEbib}
\bibliography{strings,refs}

\end{document}